\documentclass[11pt]{article}

\usepackage[numbers,sort&compress]{natbib}
\usepackage{euscript}
\usepackage{amssymb}
\usepackage{amsfonts}
\usepackage{amsbsy}
\usepackage{epsfig}
\usepackage{amsthm}
\usepackage{amscd}
\usepackage{amstext}
\usepackage{verbatim}
\usepackage{amsmath}
\usepackage{cancel}
\usepackage{capt-of}
\usepackage{empheq}

\usepackage[pdftex]{hyperref}

\def\ben{\begin{equation}}
\def\een{\end{equation}}

\def\be{\begin{equation}}
\def\ee{\end{equation}}
\def\beq{\begin{equation}}
\def\eeq{\end{equation}}
\def\ba{\begin{array}}
\def\ea{\end{array}}

\def\dalemb#1#2{{\vbox{\hrule height .#2pt
       \hbox{\vrule width.#2pt height#1pt \kern#1pt
               \vrule width.#2pt}
       \hrule height.#2pt}}}
\def\square{\mathord{\dalemb{6.8}{7}\hbox{\hskip1pt}}}
\newcommand{\bea}{\begin{eqnarray}}
\newcommand{\eea}{\end{eqnarray}}
\newcommand{\mt}[1]{\textrm{\tiny #1}}

\newcommand{\rh}{r_\mt{H}}

\newcommand{\cf}{{\cal F}}
\newcommand{\cb}{{\cal B}}
\newcommand{\ch}{{\cal H}}
\newcommand{\cn}{{\cal N}}

\begin{document}

\begin{center}

{ \Large {\bf
Notes on   shear viscosity  bound violation in anisotropic models }}

\vspace{1cm}

Xian-Hui Ge

\vspace{1cm}

{\small
{\it Department of Physics, Shanghai University,
Shanghai, 200444, China\\  }
{\it Kvali Institute for Theoretical Physics, Beijing, 100190, China}\\
E-mail: {{gexh@shu.edu.cn}}}
\vspace{1.6cm}

\end{center}

\begin{abstract}

 The shear viscosity bound violation in Einstein gravity for anisotropic black branes is discussed, with the aim of constraining the deviation of the shear viscosity-entropy density ratio  from the shear viscosity bound  using  causality  and thermodynamics analysis.
The results show that no stringent constraints can be imposed. The diffusion bound in anisotropic phases is also studied. Ultimately, it is concluded that shear viscosity violation  always occurs  in cases where the equation of motion of the metric fluctuations cannot be written in a form identical to that of the  minimally coupled massless scalar fields.
\end{abstract}
\pagebreak
\setcounter{page}{1}

\section{Introduction}

Recently, the study of transport in incoherent metals has attracted intense interest \cite{hartnoll}.
In such metals, transport is governed by the collective diffusion of energy and charge rather than by quasiparticle or momentum dissipation \cite{hartnoll}. Many  strongly correlated materials, such as 3d-transition metal-oxide compounds,
fullerenes and, most remarkably, the members of the strange metal regime of doped resistivity, exhibit a lack of a Drude peak in the optical conductivity, along with  non-motonic Hall constant temperature dependence  and thermopower \cite{brooks}.

Motivated by the anti-de Sitter/conformal field theory (Ads/CFT) correspondence, Hartnoll recently proposed a universal bound $D  \gtrsim C { \hbar v^2_{F} }/{k_B T}$ on the charge diffusivity in an incoherent metal, where $D$ is the diffusion constant, $C$ is a constant,  $v_F$ is the Fermi velocity, $T$ is the temperature,  $k_B$ is the Boltzmann constant and $\hbar$ is the Planck constant, respectively. The optical conductivity of incoherent metals can cross the Mott-Ioffe-Regel (MIR) bound, so it can be either weaker or stronger than the MIR bound. D was proposed to replace the MIR bound in bad metals and can be derived from the Kovtun-Son-Starinets (KSS) bound  $\eta/\textit{s} \geq C {\hbar}/{k_{B}}$ and the relation ${\eta}/{\textit{s}}={D T}/{v^2_{F}}$ for a vanishing chemical potential, where eta and s are the shear viscosity and entropy density, respectively  \cite{hartnoll} \footnote{In higher-derivative gravity, counterexamples to the original KSS statement have been found, but it is generally believed that a specific $C$ does exist such that a lower shear viscosity bound is always present.}.
The KSS bound is in good agreement with  relativistic heavy ion collider (RHIC) data \cite{shuryak}, and it is also well supported by the results of experiments on cold degenerated Fermi gases in the unitary limit  \cite{cao}.

However, the KSS bound is seriously challenged by the shear viscosity tensor of anisotropic, strongly coupled $\cn=4$ Super-Yang-Mills plasma derived from type-IIB supergravity \cite{rehban}. To date,
 this violation is the only such example in Einstein gravity (i.e., ${\eta}/{s}={\hbar}/{4\pi k_{B} \mathcal{H}({u_H})}$) \cite{rehban,mamo}, where $\mathcal{H}({u_H})$ is a function related to anisotropy. This is differert to higher-derivative gravity, in which the specific $C$ is stringently constrained by the fundamental causality structure of the boundary CFT. Basically, an arbitrary violation of the KSS bound would occur if $\mathcal{H}({u_H})\rightarrow \infty$, so a specific $C$ cannot exist.
In turn, $D$ would also break down \cite{pakhira1,pakhira2,dibakar2,dibakar3,pan}. Therefore, it is of fundamental importance to constrain $\ch(u_H)$ for such an anisotropic black brane.

In this paper, we first examine whether violations of such bounds can be related to any constraints on the dual gravitational side. We then examine the Hartnoll conjecture on diffusion bounds in  anisotropic black brane systems. Clear explanations for the causes and conditions of  shear viscosity bound violation are provided. A summary is presented in the final section.

\section{Charged and anisotropic axion-dilation-gravity solution }
\label{sec:incoh}

In Refs.\cite{ge,ge1}, a charged version of the anisotropic black brane solution was obtained via the five-dimensional axion-dilaton-Maxwell-Einstein bulk action reduced from type-IIB supergravity
\begin{equation}\label{action}
S=\frac{1}{2 \kappa^2} \int_{\mathcal{M}} \sqrt{-g} \Big(\mathcal{R}+12-\frac{1}{2}(\partial\phi)^{2}-\frac{1}{2}e^{2\phi}(\partial\chi)^{2}-\frac{1}{4} F_{MN}F^{MN}\Big)d^5 x+\frac{1}{2 \kappa^2} \int_{\partial\mathcal{M}}d^4 x\sqrt{-\gamma}2K,
\end{equation}
where $\kappa^2=8\pi G_5=4 \pi^2/N^2_c$ is the Newton constant, the AdS radius $L=1$, the Maxwell field strength $F_{MN}=\partial_{M} A_{N}-\partial_{N} A_{M}$,  $\chi$ is an axion field, $\gamma$ is the determinant of the induced metric on the boundary  and $K$ is the trace of the extrinsic curvature.
The black brane solution  takes the form
\begin{eqnarray}\label{metric}
&&ds^2=e^{-\frac{1}{2}\phi}r^2\Big(-\mathcal{F}\mathcal{B}dt^2+dx^2+dy^2+\ch dz^2\Big)+\frac{e^{-\frac{1}{2}\phi}dr^2}{r^2\mathcal{F}},\\
&&\chi=az,~~~A_M dx^M=A_t dt,~~~\phi=\phi(r),
\end{eqnarray}
where $A_t$ is dual to the chemical potential and a is the anisotropy parameter. The metric functions $\phi$, $\mathcal{F}$, $\mathcal{B}$, and $\mathcal{H}=e^{-\phi}$ are functions of the radial coordinate $r$ only.
The temperature $T$, the charge density $\rho$, and $s$ can be expressed as
\bea
T  &=& \sqrt{\mathcal{B}(\rh)}\bigg[\frac{\rh e^{-\frac{\phi_H}{2}}}{16\pi}\bigg(16+\frac{a^2 e^{7\frac{\phi_H}{2}}}{\rh^2}\bigg)-\frac{e^{2\phi_H}q^2 \rh}{2\pi}\bigg] ,\label{temperature} \\
s  &=&\frac{N^2_c e^{-\frac{5\phi_H}{4}}\rh^3}{2\pi}, ~~~~\rho=\frac{\sqrt{3}q \rh^3}{\kappa^2},~~~~\mu=\int^{\infty}_{\rh}dr Q\sqrt{\cb}e^{\frac{3}{4}\phi}/r^3.
\eea
The thermodynamics of this setup and its phase structure are discussed in Refs.\cite{ge,ge1}. Essentially, two types of instabilities exist: scheme-dependent and  scheme-independent. Upon holographic renormalization,
 a reference scale $\Lambda$ must be introduced.
The energy density and the pressure
transformed under the rescaling $(a,T)\rightarrow (ka,kT)$ contain an inhomogenous component caused by the presence of a non-zero conformal anomaly $\mathcal{A}={N^2_c a^4}/{48 \pi^2}$. As $F''=(\partial^2 F/\partial a^2)_{T,\mu}<0$, the system becomes unstable against infinitesimal fluctuations. Besides the scheme-dependent instability that is dependent on $\Lambda$, we uncovered scheme-independent instability by exploring the temperature-horizon radii relation and the entropy behavior. At a fixed temperature, there are two distinct branches of black brane solution with larger or smaller radii. The smaller branch is unstable
with a negative specific heat, independent of $\Lambda$.

\section{Causality}
One may naturally use the causality to constrain a. For an anisotropic fluid, the viscosity tensor $\eta_{ijkl}$ yields two shear viscosities out of five independent components \cite{landau}.
For the longitudinal mode metric fluctuation $h_{xz}(t,u,y)$ with momentum $k_y$ along $y$-direction and frequency $\omega$, we obtain the equation of motion
\bea
\partial_u(\cn^{uz}\partial_u h_z^x)-k^2_y \cn^{zy} h_z^x-\omega^2 \cn^{tz}h_z^x=0,
\eea
with the notation
\bea
\cn^{\mu \nu}=\frac{1}{2\kappa^2}g_{xx}\sqrt{-g}g^{\mu\mu}g^{\nu\nu}.
\eea
To observe the causality structure on the boundary, we simply assume $h^x_z=e^{-i\omega t+i k_y y+i k_{u}u}$. In the large momentum limit, the effective geodesic equation can be recast as
$k^{\mu}k^{\nu}g^{\rm eff}_{\mu\nu}=0$.  The effective metric can be given by
\be
ds^2_{\rm eff}=\cf\cb(-dt^2+\frac{1}{\cf\cb}dy^2)+\frac{1}{\cf}du^2.
\ee
The local speed of light is given by
\be
c^2_g=\cf\cb
\ee
and it can be expanded in terms of the standard Fefferman-Graham (FG) coordinate near the boundary
\bea
c^2_g-1=\bigg(\cb_{4}+\cf_{4}-\frac{121a^4}{576}\bigg)v^4+\mathcal{O}(v^6).
\eea
As the local speed of a graviton should be smaller than 1 (the local speed of light of the
boundary CFT), we require
\be
\bigg(\cb_{4}+\cf_{4}-\frac{121a^4}{576}\bigg)\leq 0.
\ee
In the large anisotropy limit, we calculate $\cb_{4}+\cf_{4}-{121a^4}/{576}=0$ exactly \cite{mateos}, and we obtain similar results for the transverse mode $h_{xy}$.
Therefore, no causality violation occurs for the anisotropic black brane.

\section{Thermodynamic  constraints}
Recall that the necessary and sufficient conditions for local thermodynamic stability in the canonical ensemble can be given by the specific heat at fixed charge density $c_{\rho} >0 $ and the second derivative of the free energy $F^{''}>0$. The latter condition is related to the holographic renormalization scheme. Here, we primarily study $c_\rho$,which must be positive such that
\be
c_{\rho}=\bigg(\frac{\frac{\partial E}{\partial \rh}}{\frac{\partial T}{\partial \rh}}\bigg)_{\mu,a}> 0.
\ee
For a neutral black brane, we arrive at $a^2 e^{7\phi_H/2} /\rh^2<24$ . In other words, we have ${1}/{\ch(\rh)}<({24 \rh^2}/{a^2})^{2/7}$. This means that, for a sufficiently large $a$, the
 $c_\rho$ positivity requirement can be easily satisfied. Therefore, the thermodynamic constraints do not stringently limit $a$ or ${\ch(\rh)}$. A similar result can be obtained for the non-zero chemical potential cases.

\section{Diffusion bound violation}
 Note that $a$ controls both the anisotropy and the strength of the momentum dissipation mechanism, that is, the momentum dissipation occurs more rapidly with  increased $a$. The relaxation time is found to be
 $\Gamma \propto a^2$\cite{glns}. The DC transport coefficients in this system were given in Ref.\cite{glns},where
\bea
\sigma=r_{H}e^{\frac{\phi_H}{4}}+12 r^3_H e^{-\frac{3\phi_H}{4}}\frac{q^2}{a^2},\\
\alpha=\bar{\alpha}=\frac{4 \pi Q}{a^2 e^{2\phi_H}},~~~~\bar{\kappa}=\frac{4\pi s T}{a^2 e^{2\phi_H}}.
\eea
Transport in incoherent metals is described by diffusive physics, not by momentum diffusion but rather by the diffusion of charge and energy.
The diffusion constants $D_{+}$ and $D_{-}$ are related to the transport coefficients via the Einstein relations \cite{hartnoll}
\bea
D_+ D_- & = & \frac{\sigma}{\chi} \frac{\kappa}{c_\rho} \,, \label{eq:e1} \\
D_+ + D_- & = & \frac{\sigma}{\chi} + \frac{\kappa}{c_\rho}
+ \frac{T(\zeta \sigma - \chi \alpha)^2}{c_\rho \chi^2 \sigma} \,, \label{eq:e2}
\eea
where $\sigma$, $\alpha$, and $\kappa$ are the electric, thermoelectric, and thermal conductivities, respectively. Further, $\zeta$ is the thermo-electric susceptibility and $\chi$
is the compressibility.These terms are defined as
\be
\chi=\bigg(\frac{\partial \rho}{\partial \mu}\bigg)_{T}, ~~~\zeta=\bigg(\frac{\partial \rho}{\partial T}\bigg)_{\mu},~~~c_{\rho}=T\bigg(\frac{\partial s}{\partial T}\bigg)_{\mu}-\frac{\zeta^2 T}{\chi}.
\ee
As all the physical parameters on the right-hand sides of equations (\ref{eq:e1}) and (\ref{eq:e2}) are known, we can solve for $D_+$ and $D_-$.
We are most interested in the high-temperature regime in which the localization phenomena may be irrelevant.
Considering the large temperature limit of $D_+$ and $D_-$ with vanishing chemical potential $\mu\rightarrow 0$, we obtain
\bea
\frac{k_{B}}{\hbar v^2_{F}}D_- & = &\frac{1}{\pi T}-\frac{1+4 \ln 2}{24 \pi^3 T^3}a^2-\frac{221 (\ln 2)^2-16-76 \ln2}{4608 \pi^5 T^5}a^4,\\
\frac{k_{B}}{\hbar v^2_{F}}D_+& = & \frac{4\pi T}{3a^2}+\frac{1+6\ln 2}{9 \pi T}-a^2\frac{97596-16648 \pi^2-49944 \ln 2+436167 (\ln2)^2}{922752 \pi^3 T^3},
\eea
 In the relativistic limit, we can simply use the notation $v_F=k_B=\hbar=1$.
$D_{+}$ diverges in the $T\rightarrow \infty$ limit and the same term appears in equation (11) of Ref.\cite{amoretti}. It is not easy to distinguish between the intrinsic and extrinsic contributions.
We observe that D is not saturated, although the intrinsic contributions to $D_+$ and $D_-$ at criticality are dominated by $1/T$.

\section{Discussion}
To understand the causes of shear viscosity bound violation, we must first understand how $\eta/s=1/4 \pi$ has been established. In a rotationally invariant field theory, there is only one component of $\eta$, which can be computed using Kubo's formula
\be
\eta=\lim_{\omega\rightarrow 0}\frac{1}{2\omega}\int dt d\vec{x} e^{i\omega t}\langle[T_{xy}(t,\vec{x}), T_{xy}(0,0)]\rangle,
\ee
where $T_{xy}$ is the $xy$ component of the stress-energy tensor. On the other hand, according to the gauge-gravity duality, for a graviton of frequency $\omega$  polarized in the $xy$ direction and propagating perpendicular to the black brane, the absorption cross section  is given by
\be
\Sigma_{abs}=-\frac{2\kappa^2}{\omega}{\rm Im} G^{R} (\omega)=\frac{\kappa^2}{\omega}\int dt d\vec{x} e^{i\omega t}\langle [T_{x_i x_j}(t,\vec{x}),T_{x_i x_j}(0,0)]\rangle.
\ee
Therefore, we can easily relate the shear viscosity to the graviton absorption
\be
\eta=\frac{\Sigma_{abs}(\omega=0)}{16 \pi G}.
\ee
 The relationship between the graviton $\sum_{abs}$ at low energy and the black hole entropy must now be examined. For spin-2 gravitons corresponding to $h_{xy}$, the equation of motion for $h_{x}^y$ can be expressed as $\square h_{x}^y=0$, which is indeed the equation for a minimally coupled massless scalar. The $\sum_{abs}$ of a graviton therefore has the same value as that of the scalar. There is a well-known theorem that $\sum_{abs}$ is equal to the area of the black hole horizon at low frequency, i.e., $\Sigma_{abs}(\omega=0)=A$. Using $s={A}/{4 G}$, we obtain
\be
\frac{\eta}{s}=\frac{1}{4\pi}.
\ee
For a minimally coupled massless scalar field $ h_{x}^y\equiv \phi$, the effective action is given by
\be
S=-\frac{1}{2}\int \sqrt{-g}  g^{\mu\nu} \nabla_{\mu}\phi \nabla_{\nu}\phi dt d\vec{x}.
\ee
However, for higher-derivative gravities, such as the Gauss-Bonnet gravity investigated in Refs.\cite{gb1,gb2,gb3,gb4,gb5,Sadeghi:2015vaa,dibakar1}(see also \cite{lu1,lu2}), the effective action for $h^y_x$ is expressed in momentum space as
\be
S=-\frac{1}{2}\int\frac{d \omega d^{d-1}k}{(2\pi)^d}dr \sqrt{-g}\bigg[\frac{g^{rr}(\partial_r \phi)^2}{\mathcal{Q}(r,\omega,k)}+\mathcal{P}(r,\omega,k)\phi^2\bigg].
\ee
Confirming that  the equation of motion for $\phi$ cannot be given in the same form as the massless Klein-Gordon equation is straightforward. Instead,
 the equation of motion for the tensor-type perturbation $h^y_x$ in the zero frequency and momentum limit (i.e., $\mathcal{P}=0$) can be expressed as
\be
\square h^y_x- \frac{\sqrt{-g}g^{rr}\partial_r \phi \mathcal{Q}'}{\mathcal{Q}} =0.
\ee
In this case, we obtain \cite{gb1}
\be
\frac{\eta}{s}=\frac{4 G}{\mathcal{Q}(\rh,0,0)}=\frac{1-4 \lambda}{4\pi}.
\ee
Fortunately, the Gauss-Bonnet coupling constant is strictly constrained by the causality of the boundary field theory Refs.\cite{gb1,gb2,gb3,gb4,gb5} and thus greater violation of the KSS bound is not permitted.

As regards the anisotropic black branes, $\eta$ becomes a tensor with two nontrivial components out of five. The rotational symmetry is broken and,thus, Goldstone vector bosons are generated \cite{jain}. Basically, there
are two types of metric perturbations:
\bea
&&{\rm spin~ 2: { \emph{h}_{\emph{xy}}}}\nonumber ;\\
&&{\rm spin~ 1: {\emph{h}}_{\emph{xz}}~~or~~\textit{h}_{\emph{yz}} }\nonumber.
\eea
For the spin-2 metric perturbation, $h_{xy}$ still obeys an equation of motion identical to that of a minimally coupled massless scalar. For the spin-1 metric perturbation, $h_z^x$, the equation of motion is
$\nabla_{\mu} f^{\mu\nu}+f^{\mu\nu}{\nabla_{\mu}g_{xx}}/{g_{xx}} =0$, with $f_{\mu\nu}=\partial_{\mu}a_{\nu}-\partial_{\nu}a_{\mu}$ and $a_z=h_z^x$. The $\eta$ corresponding to the spin-$1$ component may behave similarly to the conductivity \cite{gsw}. Moreover, for vector absorption, $\sum_{abs}$ approaches zero as $\omega\rightarrow 0$.  We cannot expect that $\sum_{abs}$ remains equal to the black brane horizon area in such a case. As discussed in the previous sections, the $\eta$ of the spin-$1$ component can violate the KSS bound in a parametric manner. That is, $\eta/s$ can be as small as possible with no strong constraints from causality and thermodynamical instability.

\section{Summary}
In summary, we have studied the shear viscosity and diffusion bounds in anisotropic black branes and provided an explicit explanation of the causes of shear viscosity violation.  It is clear that the corresponding metric perturbation cannot be expressed in the form of a minimally coupled massless Klein-Gordon equation when the KSS bound is violated. We failed to obtain a minimal KSS bound value by considering the causality and thermodynamic  instability constraints. This differs significantly from the Gauss-Bonnet gravity, where the Gauss-Bonnet coupling constant is strictly constrained by the causality problem of the boundary field theory. We also examined the diffusion bound in our model. As the KSS bound is violated, it is not surprising that the diffusion bound is not saturated.

On the other hand, based on the dynamical mean field theory (DMFT), Pakhira and McKenzie have found that both the KSS and Hartnoll bounds are violated in a single-band Hubbard model \cite{pakhira1,pakhira2}. For the shear viscosity to entropy density ratio for an electron fluid, which is described by a single-band Hubbard model at half filling, the KSS bound was found to be strongly violated in the bad-metal regime for parameters appropriate to lattice electronic systems, such as cuprates and organic charge-transfer salts. Moreover, the diffusivity bound was shown to be violated by several orders of magnitude in the incoherent transport regime. It would be worth exploring whether these observations can be related to the results obtained here. On a more theoretical note, it would be interesting  to find an alternative bound for such anisotropic systems in future work \cite{gsw}. For studies on  the universal features of the Langevin diffusion coefficients, one may refer to \cite{Giataganas12,Giataganas13,Giataganas14}.

\section*{Acknowledgements}

The author would like to thank J. X. Lu, Hong L$\ddot{u}$, Yu Tian,  Qun Wang   and Shao-Feng Wu for helpful discussions at different stages of this work.  The author was partly supported by NSFC, China (No.11375110).

\end{document}